\documentclass[preprint,nobibnotes,showpacs,showkeys]{revtex4}
\usepackage{amssymb}
\usepackage{amsfonts}
\usepackage[dvips]{color}
\usepackage{newlfont}
\usepackage{graphicx}
\usepackage{amsmath}
\usepackage[latin1]{inputenc}
\usepackage{float}

\begin{document}

\title{ Complementarity and Classical Limit of Quantum Mechanics: Energy Measurement aspects }
\author{Ad\'elcio C. Oliveira}
\affiliation{Departamento de F\'isica e Matem\'atica, Universidade Federal de S\~ao
Jo\~ao Del Rei, Ouro Branco, 36420 000, Minas Gerais , Brazil}
\email{adelcio@ufsj.edu.br}
\author{ Zolacir T. Oliveira Junior}
\affiliation{Departamento de Ci\^encias Exatas e Tecnol\'ogicas, Universidade Estadual de
Santa Cruz, Ilh\'eus, 45662 000, Bahia, Brazil}
\author{Nestor S. Correia}
\affiliation{Departamento de Ci\^encias Exatas e Tecnol\'ogicas, Universidade Estadual de
Santa Cruz, Ilh\'eus, 45662 000, Bahia, Brazil}

\begin{abstract}
In the present contribution we discuss the role of experimental limitations
in the classical limit problem. We studied some simple models and found that
Quantum Mechanics does not re-produce classical mechanical predictions,
unless we consider the experimental limitations ruled by uncertainty
principle. We have shown that the discrete nature of energy levels of
integrable systems can be accessed by classical measurements. We have
defined a precise limit for this procedure. It may be used as a tool to
define the classical limit as far as the discrete spectra of integrable
systems are concerned. If a diffusive environment is considered, we conclude
that the "lifetime" of discreteness is approximately $1/\kappa$ ($\kappa$ is
the diffusion constant), thus it was possible to relate the classical limit
of a spectra with the action of an environment and experimental resolution.
\end{abstract}

\pacs{03.65.Yz, 42.50.Lc, 03.67.Bg, 42.50.Lc,
42.50.Pq}
\keywords{ complementarity, classical limit, energy spectrum, correspondence
principle, experimental accuracy, uncertainty principle}
\maketitle

\section{Introduction}


Since the early twentieth century, the predictions of quantum mechanics were
strange to most people and almost always seemed to contradict our common
sense. There were no conclusive explanation for the disappearance of quantum
effects in the macroscopic regime. Which of the quantum features we don't
see in our day life experience? One of the most disconcerting one is the
wave-particle duality. This is in fact the first quantum feature that has a
strong argument explaining its disappearance in the macroscopic world. The
solution is in de Broglie expression for the wavelength. The length of the
wave interference fringes becomes insignificant as we increase the classical
action of the system. There are many other quantum features that are
unobservable in the macroscopic regime, such as entanglement and
discreteness of energy spectrum. Other important open question is about the
classical regime. Does quantum mechanics reproduce the observed results of
macroscopic experiments? The first try to answer this question is attributed
to Bohr, the correspondence principle. Bohr's correspondence principle \cite%
{Bohr}:

"Indeed, as adequate as the quantum postulates are in the phenomenological
description of the atomic reactions, as indispensable are the basic concepts
of mechanics and electrodynamics for the specification of atomic structures
and for the definition of fundamental properties of the agencies with which
they react. Far from being a temporary compromise in this dilemma, the
recourse to essentially statistical considerations is our only conceivable
means of arriving at a generalization of the customary way of description
sufficiently wide to account for the features of individuality expressed by
the quantum postulates and reducing to classical theory in the limiting case
where all actions involved in the analysis of the phenomena are large
compared with a single quantum. In the search for the formulation of such a
generalization, our only guide has just been the so called correspondence
argument, which gives expression for the exigency of upholding the use of
classical concepts to the largest possible extent compatible with the
quantum postulates."

Bohr also stated the postulate known as selection rule:

"This peculiar relation suggests a general law for the occurrence of
transitions between stationary states. Thus we shall assume that even when
the quantum numbers are small the possibility of transition between two
stationary states is connected with the presence of a certain harmonic
component in the motion of the system \cite{Bohr1920}. "

Bohr's correspondence principle and his selection rule were later
interpreted as \cite{Home}:

\begin{enumerate}
\item ``Quantum mechanical predictions must correspond to classical ones in
the limit of large quantum numbers.''

\item \textquotedblleft A selection rule must be valid for all possible
quantum numbers. Thus, a selection rule must be valid for classical and
quantum regimes.\textquotedblright\
\end{enumerate}

The Correspondence Principle of Quantum Mechanics is a widely used tool to
study quantum to classical transition problem since its conception by Bohr
\cite{Bohr1920,Bohr}. Nevertheless, Quantum and Classical Mechanics are
incommensurable theories because their basic ontologies are different. The
basic entities of Classical Mechanics (CM) are coordinates and velocities in
configuration spaces. Quantum Mechanics (QM) deals with state vectors and
operators in Hilbert space \cite{Laloe}. Since both theories give acceptable
descriptions of different experiments in limiting situations there should be
a way of translating their predictions from one to the other, in going from
QM to CM this is the so called classical limit problem \cite{Zur1996}. There
are also attempts to explain experimental results usually in quantum domain
within the framework of CM \cite{Rubin,Luca,Figueiredo, Gesil}.

In the core of the Correspondence Principle there are two ``beliefs''\ that
has influenced the study of the classical limit problem. The first one is
that classical mechanics is able to reproduce macroscopic experimental
observations within an arbitrary precision in a such way that its
predictions can be used for comparison with the quantum description of the
experiment. The second one is that ``typical quantum objects''\ and
``typical classical objects'' are subjected to the same conditions. Although
the confirmation of these ``beliefs'' may be desirable, this is not
necessary \footnote{%
Actually, it is largely accepted that decoherence, which is an inherently
quantum process \cite{Zur2003} and does not exist in classical mechanics,
works as a selection rule for the transition from quantum to classical.} nor
sufficient. We know that an experimental result, usually in classical
domain, can be considered as a quantum one \cite{Yu02, Inomata05}. This
problem can be extended to a more general problem of comparing two theories.
Suppose we have two theories \textsl{A} and \textsl{B}, and that it is
common sense that theory \textsl{A} has a validity domain that includes the
domain of \textsl{B}, i.e., \textsl{A} $\supset $ \textsl{B}. Thus, it is
expected that theory \textsl{A} reproduces all the results for \textsl{B}
and it should also not be in contradiction with theory \textsl{B} in its
domain. A typical example is special relativity, where we adopt slow
velocity limit to yield Newtonian regime. Strictly speaking about classical
mechanics, some points are relevant to our discussion:

\begin{enumerate}
\item Classical mechanics is a very successful theory, but it does not mean
that there is no uncertainty within its predictions. Inaccuracies are
unavoidable, but in classical domain Newtonian dynamics is in agreement with
the observations.

\item It would be very interesting to obtain classical mechanics from
quantum mechanics in terms of a limit, such as high energy or mass. Although
that, the parameter to be used should not be a universal constant such as $%
\hbar $. For practical reasons it should be a variable that can be
manipulated in a laboratory.

\item There is not a known experimental criteria of general validity which
determines if a quantum state can be considered classical in all aspects or
all possible measurements.
\end{enumerate}

The above comments take us to philosophical questions, such as what is the
objective of a physical theory? What do we need to build a useful theory?
Which criteria one should use to compare two theories? It is common sense
that a good physical theory must reproduce the experimental results within
the error bar. It would be more acceptable if it is able to predict unknown
phenomena. A physical theory must include mathematical entities which the
physicist should associate with the ones observed in laboratory. With the
help of such mathematical entities, one should be able to predict the time
evolution of the observed physical entities. The last question is a subject
of a stronger debate \cite{Zur1996, Ball98, Ball01, Oliveira02, Dalvit,
Wiebe, Oliveira06, Faria07, Davidov, Bosco, adelcio2012, Gabi, renato2007} .
If it was always possible to measure the quantum state, it would be a most
precise separability criteria. But in fact we know how to measure Wigner
functions only in some specific situations \cite{Lutter1,Lutter2,
Davidov,Banaszek,Mukamel,Kenfack,Toscano06}. Let us go back to the general
problem of theories \textsl{A} and \textsl{B}. Suppose that an
experimentalist makes an experiment in \textsl{B} domain, and the result is
compatible, within experimental error bar, with theory \textsl{B}. If theory
\textsl{A} also gives a good description of the experiment and its
prediction is in accordance with the results, we can thus say that for this
specific experiment both theories are satisfactory, meaning that they are
equivalent for this case. If $Im(A)\bigcap $Im(B)$\neq \emptyset $, it is
possible to have such an experimental result compatible with theories
\textsl{A} and \textsl{B}: in this sense, the theories are not divergent.
Due to the experimental errors \cite{Rajeev, adelcio2012}, or inaccuracies
\footnote{%
In this context error means experimental resolution.}, that are unavoidable
and inherent to quantum theory\footnote{%
We should remark that there is not yet a quantum measurement error theory,
although there is some effort in this direction \cite{Gardiner,Rajeev,
adelcio2012}.}, a point in domain should be represented as a volume in its
image. If all possible realizable experiments in \textsl{B} domain have
their image in the intersection of theories image we can thus say that
theory \textsl{A} successfully replaces \textsl{B}.

The investigation of the correspondence principle employing a quartic
oscillator coupled to a diffusive phase reservoir and taking into account
the limited experimental resolution is found in Ref. \cite%
{renato2007,adelcio2012}. They studied the system dynamics in perspective of
an observable, the position or momentum. \bigskip

In ref. \cite{adelcio2012} they show that the problem of the
quantum-to-classical transition is close related to decoherence and
inaccuracies of measurements on the correspondence of the two descriptions.
They claimed that isolated systems and measurements with unlimited accuracy
are mere idealizations.

In this contribution, the main goal is to investigate the role of
experimental errors and \emph{of environment} in the quantum-classical
transition problem. We focuss on the spectra discreteness. We propose a
heuristic procedure (not deduced mathematically within the theory) to use
spectroscopic information from model Hamiltonians and time energy Heisenberg
relations in order to decide whether a quantum system can be described by
CM. The quantum behavior is characterized by the discreteness of energy
spectra. We are considering a gedankenexperiment, where the experimentalist does not know Quantum Mechanics but tries obtain the energy spectrum. Will he  be able to conclude that it is discrete?

\bigskip

\section{\protect\bigskip The large quantum numbers limit}

One controversial subject in quantum mechanics is the time-energy
uncertainty relation ($\Delta t.\Delta E\geq \frac{\hbar }{2}$) and it is
not our aim to bring this discussion to the present work. Nevertheless it is
well known that time and frequency are conjugated variables in a pair of
Fourier transforms in classical physics, the duration of a signal and the
respective frequency are subjected to an unsharpness relation ($\Delta
t.\Delta \omega \geq \frac{1}{2}$) that is classical indeed. Frequency, in
quantum theory, is another way of speaking about energy. An uncertainty
relation between time and energy must be seriously considered, studied and
interpreted, although there are objections due to the fact that time is not
associated to a dynamical operator canonically conjugated to the
hamiltonian. An exhaustive examination of this matter is made by Peres in
his book \cite{Peres 2002}. On the other hand we have to assume that QM is
not an objective description of physical reality. It only predicts the
probability of occurrence of \textit{stochastic macroscopic events},
following specified preparation procedures. These predictions are based on
conceptions that are given to know indirectly - like all the microscopic
elements as electrons, photons, etc. They have no substance at all, they are
just perceived or understood by their interactions with measurement devices.
We want to avoid these interpretational intricacies of QM by stating the
problem in another way: how to describe the aim of measurement or what \ has
actually to be done in a measurement process?

The main stream of our proposal is not face the discussion of the
time-energy uncertainty relation, we face the problem of to decide under
what conditions a system may be considered as Classical or Quantum. It is in
this sense that we tackle the problem of dealing with the physical reality:
how can one measure and what is in fact measured. Thus, we use the product
of energy differences between neighbour levels by the corresponding
classical period differences to compare with the time-energy uncertainty
relation to classify a system as classical or quantum. If this product
fulfills the time-energy uncertainty relation it is quantum otherwise it is
classical.

To do this we add an element connected to this subject that concerns
integrable systems: the decision wether a system is classical or quantum
depends on the experimental apparatus which are essentially classical. One
undoubtedly quantum feature is the discreteness of at least part of the
energy spectrum. The essential idea here is that if one tries to measure the
energy of the system in question using classical canonical pairs $q$, $p$,
the information about the quantum nature of the particle will be lost. In
spite of this fact, as we show in what follows, the function
\begin{equation}
y(n)=\left\vert \Delta E_{n}\Delta \tau _{n}\right\vert   \label{yfunction}
\end{equation}%
where $\Delta E_{n}=\left( E_{n}-E_{n-1}\right) /2$ is merely the energy
difference between two neighbour levels, (it is the maximum uncertainty in
energy for the state $a\left\vert E_{n}\right\rangle +b\left\vert
E_{n-1}\right\rangle $  ) and $\Delta \tau _{n}=\left( \tau _{n}-\tau
_{n-1}\right) /2,$ with $\tau _{n}$ being the classical period associated to
the energy $E_{n}.$ If the experiment has an accuracy $\delta t$ it limits
the period measurement precision, in real systems it is desirable that $%
\delta t<<\Delta \tau $, we assume that they are of same order, $\Delta \tau
\approx $ $\delta t.$

The $y(n)$ function can be heuristically justified if we consider the
Bohr-Sommerfeld quantization rule for periodic systems, that states
\begin{equation}
I=\oint pdq=2\pi \hbar n
\end{equation}%
thus we have%
\begin{equation}
\left\langle K\right\rangle =\frac{\pi n\hbar }{\tau },  \label{ktau}
\end{equation}%
where $\left\langle K\right\rangle $ is the the mean kinetic energy, $\tau $
being the classical period associated with $K$ and $n$ is a quantum number.
Observing that \cite{Darrigol}
\begin{equation}
\delta I=\oint \delta Hdt
\end{equation}%
Consider two neighboring periodic motions of the same system. Then H is a
constant, and we have

\bigskip
\begin{equation}
\delta I=\tau \delta H.  \label{deltaH}
\end{equation}%
Note that (\ref{deltaH}) determines that the energy of the system depends
only on I, then Bohr's quantiztion rule determines the energy of the system,
for more detail see \cite{Darrigol}.

The Bohr-Sommerfeld quantization rule makes a direct conection between
kinetics energy and classical period, it was also demonstrated \cite%
{adelcioJMP} that, for integrable systems, is possible to reconstruct
semiclassicaly the quantum state using classical dynamics. Since you know
the Wigner function of the system it is possible to infer the period related
to the state \cite{adelcioJMP, Oliveira01}. \bigskip Alternatively, we can
use the semiclassical quantization rule \cite{Berryi, Stockmann} for
integrable systems, we have a similar result \cite{Stockmann, Wisniacki,
Novaes}.

This function (\ref{yfunction}) is stated for a more clear definition of a
classical limit. As we will show, the fact that one will only be able to see
a continuum of energies does not mean that the classical limit has been
reached. It only reveals that we have become \textquotedblleft
myope\textquotedblright\ to see the nature of the spectrum.

In what follows, we will consider a simple one dimensional systems with
discrete spectra. Let us assume that a state has been prepared in the
following way
\begin{equation}
\left\vert \Psi (0)\right\rangle =a\left\vert E_{n}\right\rangle
+b\left\vert E_{n-1}\right\rangle  \label{1}
\end{equation}%
where $\left\{ \left\vert E_{n}\right\rangle \right\} $ are eigenstates of
the hamiltonian $H$
\begin{equation}
\hat{H}\left\vert E_{n}\right\rangle =E_{n}\left\vert E_{n}\right\rangle
\label{2}
\end{equation}%
The variance in energy difference, for any time, of this state is given by
\begin{equation}
\Delta E=\left\vert a\right\vert \left\vert b\right\vert \left[ E_{n}-E_{n-1}%
\right]
\end{equation}%
Using the normalization condition $\left\vert a\right\vert ^{2}+\left\vert
b\right\vert ^{2}=1$, it is easy to check that the maximum for this energy
difference is achieved for $\left\vert a\right\vert =\left\vert b\right\vert
=1/\sqrt{2}$, and is given by

\begin{equation}
\Delta E(n)=1/2\left\vert E_{n}-E_{n-1}\right\vert .
\end{equation}

How do we define the classical limit in this simple case? Let us assume the
following $\displaystyle\lim_{n\rightarrow \infty }\frac{\Delta E_n}{E_n}=0$.

The above expression means that as the energy increases $(E_{n})$, the
uncertainty in energy of the state $\left| \Psi (t)\right\rangle $ becomes
negligible as compared to $E_{n}.$ Let us discuss a few examples.

\bigskip \emph{Harmonic Oscillator}

For the harmonic oscillator we have

\begin{equation}
\Delta E=\frac{1}{2}\hbar \omega \Longrightarrow \frac{\Delta E}{E_{n}}=%
\frac{1/2}{(n+1/2)}  \label{delta e oscilador harmonico}
\end{equation}

and we clearly have $\displaystyle\lim_{n\rightarrow \infty }\frac{\Delta E_n%
}{E_n}=0$.

\bigskip \emph{Particle in a box}

In this case we have
\begin{equation}
E_{n}=\hbar ^{2}\frac{n^{2}\pi ^{2}}{2ma^{2}},
\end{equation}%
where $n=1,2,3,dots$, $m$ is the mass of the particle and $a$ is the box
width. Then we have
\begin{equation}
\Delta E_{n}=\hbar ^{2}/4\left\vert \frac{(2n-1)\pi ^{2}}{ma^{2}}\right\vert
.  \label{Delta E caixa}
\end{equation}%
Once again we have $\lim_{n\rightarrow \infty }{\displaystyle{\frac{\Delta
E_{n}}{E_{n}}}}\rightarrow 0.$

The fact that this limit is zero is the argument usually found in textbooks
in order to define the classical limit\cite{Home,Cohen,Eisberg}. Of course
it is true that when one considers the high energy limit, it becomes
increasingly difficult to obtain good experimental resolution. However, as
we show below, this limit does not necessarily imply that it is impossible
to obtain the needed resolution. As we know, quantization has been observed
in some macroscopic systems like superconducting Josephson junctions \cite%
{Richard}. Besides, one can use something analogous to the Heisenberg
microscope \cite{Quantummesurement} in order to obtain the velocity of the
particle. In the case of a particle in a box, discussed above, once we know
the velocity, the energy is easily obtained. The measurement of the velocity
implies a perturbation of the position which depends on the characteristics
of the apparatus and in principle, is independent of the quantum number $n$.
Thus, if one uses the adequate measurement the discrete character of a
spectrum can be verified.

\section{Classical measurement of the Energy}

When one deals with realistic systems, say, atoms, the spectrum is obtained
from the emitted or absorbed electromagnetic radiation. A completely
different approach is used for macroscopic systems. Within CM, for closed
systems, the energy is a function of position and momentum $E=f(p,q).$ So,
from now on, we will call Classical Measurement of energy every process
which makes use of the relation $E=f(p,q)$ in order to obtain the energy of
a given system, classical or quantum. As an example, we may look at the
particle in a box again. In this case, the energy is a direct function of
the velocity, so that once we determine de velocity, the energy will be
defined. In practice, one may measure the time it takes for $2s$ inversions
in the momentum and so determine the period of the motion . Classically, the
period is given by
\begin{equation}
\tau =\frac{2a}{v}=a\sqrt{\frac{2m}{E}.}
\end{equation}%
In the above expression, we are considering the correspondent classical
period for a specific energy eigenvalue. Since the energy levels are
determined by QM, regardless of the energy scale one is talking about, the
allowed values of the period are
\begin{equation}
\tau _{n}=\frac{2a^{2}m}{\hbar n\pi }.
\end{equation}%
So, the difference in period for two quantum neighboring levels is
\begin{equation}
\Delta \tau =\frac{\tau _{n}-\tau _{n-1}}{2}=-\frac{a^{2}m}{\hbar \pi }\frac{%
1}{\left( n-1\right) n}.  \label{delta tau}
\end{equation}

From the above expression one can see that it becomes increasingly difficult
to distinguish two higher adjacent energy levels by this method. Let us take
a look at the product $\left\vert \Delta E_{n}\Delta \tau _{n}\right\vert $
where $\Delta E_{n}=\left( E_{n}-E_{n-1}\right) /2$ and $\Delta \tau
_{n}=\left( \tau _{n}-\tau _{n-1}\right) /2.$ $\tau _{n}$ is the classical
period associated with the energy $E_{n}.$ It is interesting to observe the
behavior of the function
\begin{equation}
y(n)=\left\vert \Delta E_{n}\Delta \tau _{n}\right\vert .
\end{equation}%
For the case in question,
\begin{equation}
y(n)=\frac{\hbar \pi }{4}\frac{(2n-1)}{\left( n-1\right) n}.
\end{equation}%
If we take $n=4,$ we find $y(4)<\hbar /2$, which means that we won't have
enough precision to verify whether the spectrum is discrete or continuous,
since Quantum Mechanics forbids such precision\footnote[5]{\textsc{The
time-energy uncertainty relation} $\Delta E\Delta t\geq \hbar /2$ is a first
principle limitation and has nothing to do with experimental errors. A
deeper discussion on the subject can be found in Refs. \cite%
{Appleby,Rajeev,Quantummesurement}.}. Here, we are just using the fact that
the time-energy uncertainty relation has to be respected since we are
considering that Quantum Mechanics must prevent Classical Mechanics. Also it
is easy to see that $\displaystyle\lim_{n\rightarrow \infty }y(n)=0.$

In the more realistic case of a hydrogenoid atom we have
\begin{equation}
E_{n}=-\frac{\mu Z^{2}e^{4}}{2\hbar ^{2}n^{2}}  \label{E 1/r}
\end{equation}%
where $Ze$ is the total charge interacting with the electron of charge $e$,
and $\mu $ is the reduced mass of the system. As opposed to the previous
case, the energy levels become closer as $n$ grows. Thus, the question is:
\textit{from which $n$ can we say that the spectrum is continuous }from the
point of view of Classical Mechanics? In this case
\begin{equation}
\Delta E=\frac{\mu Z^{2}e^{4}}{4\hbar ^{2}}\left\vert \frac{2n-1}{%
n^{2}(n-1)^{2}}\right\vert ,  \label{Delta E 1/r}
\end{equation}%
\begin{equation}
\Delta T=\frac{\pi \hbar ^{3}}{Z^{2}e^{4}\mu }\left[ 3n^{2}-3n+1\right] ,
\label{deta T 1/r}
\end{equation}%
and
\begin{equation}
y(n)=\frac{\pi \hbar \left( 2n-1\right) \left[ 3n^{2}-3n+1\right] }{%
4n^{2}(n-1)^{2}}.
\end{equation}%
Now, for the $1/r$ potential, we have $y(n)<\hbar /2$ for $n=9$, so one
won't be able to observe the quantization for $n$ bigger than 9 while using
classical measurement of the energy.

Other interesting case is the Morse Potential, frequently used to describe
the spectra of molecules \cite{Oliveira01} . The Morse potential is defined
as
\begin{equation}
U(x)=D\left( e^{-2\alpha x}-2e^{-\alpha x}\right) ,
\end{equation}%
where $D$ and $\alpha $ are constants experimentally determined. For $s$
waves, i.e., the orbital angular momentum is zero, we have

\begin{equation}
E(n)=-D+\hbar \omega \left[ \left( n+1/2\right) -\frac{1}{\zeta }\left(
n+1/2\right) ^{2}\right] .
\end{equation}%
$\zeta $ is also experimentally determined. The classical period
corresponding to each $E(n)$ is given by
\begin{equation}
\tau (n)=2\pi \sqrt{\frac{MRo^{2}}{2\left\vert E(n)\right\vert \alpha ^{2}}}
\end{equation}%
where $Ro$ is a function of $\alpha $ and $\zeta .$ It is easy to verify
that, for the hydrogen molecule parameters, $y(n)<\hbar /2$ for all possible
$n$. Since the Morse potential is quasi-harmonic for low energies we observe
that it is not possible to distinguish neighboring discrete states with a
classical measurement. From this example, we may conclude that any potential
that is approximately harmonic have no assessable discrete spectrum through
a classical measurement of the energy.

\subsection{Harmonic Oscillator}

For a harmonic oscillator, $E=\frac{p^{2}}{2m}+\frac{1}{2}kq^{2}$\ , its
period is $\tau =\frac{2\pi }{\omega }$ where $\omega =\sqrt{\frac{k}{m}}$
which is energy independent, thus we conclude that it can not be used to
characterize the spectrum. Another way of classically determining the energy
can be obtained by measuring q and p at same time, therefore classical
energy uncertain is given by
\begin{equation}
\delta E=\frac{p}{m}\delta p+kq\delta q
\end{equation}

Without lost of generality, we choose $\delta p=\sqrt{\frac{m\hbar \omega }{2%
}}a$ and $\delta q=\sqrt{\frac{\hbar }{2m\omega }}a$ then we obtain the
relation
\begin{equation}
\delta p\delta q=a^{2}\hbar /2.
\end{equation}

In case of a=1 we have the minimum uncertain defined under Hisenberg
relation, in general we have
\begin{equation}
\delta E=\left[ |p|\sqrt{\frac{\hbar \omega }{2m}}+k|q|\sqrt{\frac{\hbar }{%
2m\omega }}\right] a
\end{equation}

then
\begin{equation}
\delta E^{2}=\frac{2}{\hbar }\left[ p\sqrt{\frac{\hbar \omega }{2m}}+kq\sqrt{%
\frac{\hbar }{2m\omega }}\right] ^{2}\delta p\delta q
\end{equation}

From (\ref{delta e oscilador harmonico}) we have

\begin{equation}
\delta E=\frac{1}{2}\hbar \omega
\end{equation}%
thus we find
\begin{equation}
\delta p\delta q=\frac{\hbar ^{3}\omega ^{2}}{4\left[ |p|\sqrt{\frac{\hbar
\omega }{2m}}+k|q|\sqrt{\frac{\hbar }{2m\omega }}\right] ^{2}}.
\end{equation}%
Note that
\begin{equation}
\left[ |p|\sqrt{\frac{\hbar \omega }{2m}}+k|q|\sqrt{\frac{\hbar }{2m\omega }}%
\right] ^{2}=\left( \frac{p^{2}}{2m}+\frac{1}{2}kq^{2}+|pq|\omega \right)
\hbar \omega ,
\end{equation}

if we consider the energy as $E=\hbar\omega(n+1/2)$, we find
\begin{equation}
\delta p\delta q=\frac{\hbar^{2}}{4\left( \hbar(n+1/2)+|pq|\right)}
\label{dqdpoh}
\end{equation}

From (\ref{dqdpoh}) we see that we need an experimental resolution that is
avoided by Quantum Mechanics. Thus we can say that his spectrum can not be
resolved by Classical Energy Measurement.

\section{Neighbors Distinguishability}

Let us consider a quartic oscillator model. Its Hamiltonian is
\begin{equation}
\hat{H}_{0}=\hbar \omega \hat{a}^{\dagger }\hat{a}+\lambda \hbar ^{2}(\hat{a}%
^{\dagger }\hat{a})^{2}  \label{Ho quartico}
\end{equation}%
where $\hat{a}$ and $\hat{a}^{\dagger }$ are destruction and creation
operators of the harmonic oscillator, $\omega $ is the natural frequency of
the main oscillator, $\lambda $ is the non-linear strength constant. The
system has as energy levels
\begin{equation}
E(n)=\hbar \omega n+\lambda \hbar ^{2}n^{2}.  \label{E quartico}
\end{equation}

Its classical counterpart, with an energy given by (\ref{E quartico}) has a
period given by

\begin{equation}
\tau =\frac{2\pi }{\omega +2\lambda \hbar n},  \label{tau quartico}
\end{equation}%
then we have
\begin{equation}
y(n)=\frac{\pi \lambda \hbar ^{2}\left[ \omega +\lambda \hbar (2n-1)\right]
}{\left[ \omega +2\lambda \hbar (n-1)\right] \left[ \omega +2\lambda \hbar n)%
\right] }.  \label{y(n) quartico}
\end{equation}

The case $\lambda \hbar \ll \omega $ we obtain harmonic potential, as
expected, and for $\lambda \hbar \gg \omega $ we obtain the infinity well
result:%
\begin{eqnarray}
\lim_{\lambda \hbar /\omega \rightarrow 0}y(n) &\approx &\frac{\pi \lambda
\hbar ^{2}}{\omega }, \\
\lim_{\lambda \hbar /\omega \rightarrow \infty }y(n) &\approx &\frac{\pi
\hbar (2n-1)}{4(n-1)n}.
\end{eqnarray}

Considering $\lambda \hbar \gg \omega ,$ in principle we can access the
discreet spectra for $n<4$.

Up to now, we didn't consider the environment action. Real systems are never
isolated, thus in a real experiment we can not ignore its action. Due to the
impossibility of modeling the system and its surrounding from first
principles, we are forced to treat the environment as a collection of
harmonic oscillators, as usual \cite{Isar}. We use the model of the quartic
oscillator \cite%
{Oliveira02,Oliveira06,renato2006,renato2007,Faria07,Greiner,Agarwal}
coupled to a bath of harmonic oscillators. This model is described by the
hamiltonian
\begin{equation}
\hat{H}_{q}=\hat{H}_{0}+\hat{H}_{int}+\hat{H}_{A}
\end{equation}%
where
\begin{equation}
\hat{H}_{0}=\hbar \omega \hat{a}^{\dagger }\hat{a}+\lambda \hbar ^{2}(\hat{a}%
^{\dagger }\hat{a})^{2}
\end{equation}%
is the hamiltonian that governs the free evolution of the system of
interest,
\begin{equation}
\hat{H}_{A}=\sum_{k}\hbar \omega _{k}\hat{b}_{k}^{\dagger }\hat{b}_{k}
\end{equation}%
is the hamiltonian that governs the free evolution of the set of
environmental harmonic oscillators, and
\begin{equation}
\hat{H}_{int}=\sum_{k}\hbar g_{k}(\hat{b}_{k}\hat{a}^{\dagger }+\hat{b}%
_{k}^{\dagger }\hat{a})
\end{equation}%
is the interaction hamiltonian between the system of interest and the
environment. In the above equations, $\omega _{k}$ is the natural frequency
of the $k$-th oscillator of the bath, and $g_{k}$ is the coupling constant
between the $k$-th bath oscillator and the system. For an initial fock state
$\left\vert b\right\rangle ,$ the density matrix evolution is given by
\begin{align}
\rho (t,b)& =\sum_{l=0}^{\infty }\sum_{p=0}^{b}\frac{b!(p+l)!}{\left(
p!\right) ^{2}!l!(b-p)!}  \label{ro(t) para fock} \\
& \times \gamma ^{b+l-p}(0,t)\zeta ^{2p+1}(0,t)\left\vert p+l\right\rangle
\left\langle p+l\right\vert .  \notag
\end{align}

\bigskip here $\left\vert n\right\rangle $ are fock states and
\begin{equation}
\gamma (n,t)=\frac{2\kappa \sinh (\Delta t)}{\Delta \cosh (\Delta
t)+(i\lambda n+2\kappa )\sinh (\Delta t)}
\end{equation}%
\begin{equation}
\zeta (n,t)=\frac{\Delta }{\Delta \cosh (\Delta t)+(i\lambda n+2\kappa
)\sinh (\Delta t)}
\end{equation}%
where $\Delta =\sqrt{(i\lambda n+2\kappa )^{2}-4\kappa ^{2}}$

\bigskip and $\kappa$ is the diffusion constant, see \cite%
{Oliveira06,Faria07}.

This equation (\ref{ro(t) para fock}) indicates that an initial fock state
turns to a statistical mix state in a short time ruled by $\kappa .$ Thus,
if we prepare the system in an initial fock state $\left\vert b\right\rangle
,$ after a time t, we can measure the system in a different fock state. If
we wish to distinguish neighboring states we may use the fidelity. The time
evolution of fidelity between $\left\vert b\right\rangle $ and $\left\vert
b-1\right\rangle $,is given by

\begin{equation}
F(b,t)=Tr\left[ \rho (t,b)\rho (t,b-1)\right] .
\end{equation}%
After some straightfort algebraic manipulations we obtain
\begin{align}
F(b,t)& =\sum_{l=0}^{\infty }\sum_{p=0}^{b}\sum_{p^{\prime }=0}^{\min
(b-1,p+l)}\{\frac{b\left( (b-1)!\right) ^{2}\left( (p+l)!\right) ^{2}}{%
\left( p^{\prime }!\right) ^{2}\left( p!\right) ^{2}!l!(b-p)!}\frac{1}{%
\left( p+l-p^{\prime }\right) !(b-p^{\prime }-1)!}  \label{Fidelidade(t)} \\
& \times \gamma ^{2b+2l-2p^{\prime }-1}(0,t)\zeta ^{2(p+p^{\prime
})+2}(0,t)\}.  \notag
\end{align}

\begin{figure}[ht]
\vspace{-0cm} \includegraphics[scale=.8]{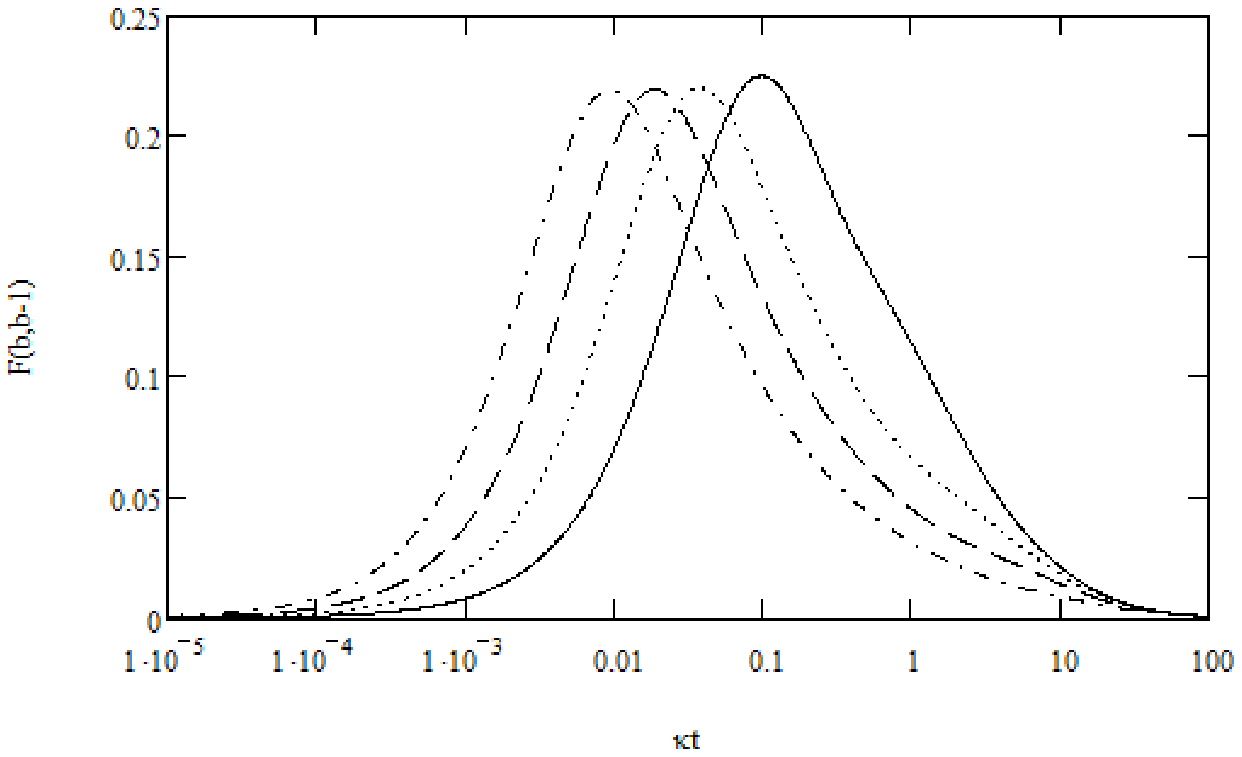} \vspace{-0cm}
\caption{Full thin line we see the fidelity decay , $F(b,t)$, for $b=1$,
dotted line $b=5$, dashed line $b=10$, dash dotted line $b=15$, the x axis
corresponds to $\protect\kappa t$ in a log scale.}
\label{fid}
\end{figure}

In figure (\ref{fid}) we see $F(b,t)$ for b=1,2,3,4 . Even for short time, $%
F(b,t)$ is different from zero, thus the gedankenexperiment of classical
energy measurement has to be performed in a short time, what in practice
turns it almost impossible since one needs a long time experiment for an
accurate energy measurement .

\begin{figure}[th]
\vspace{-0cm} \includegraphics[scale=.8]{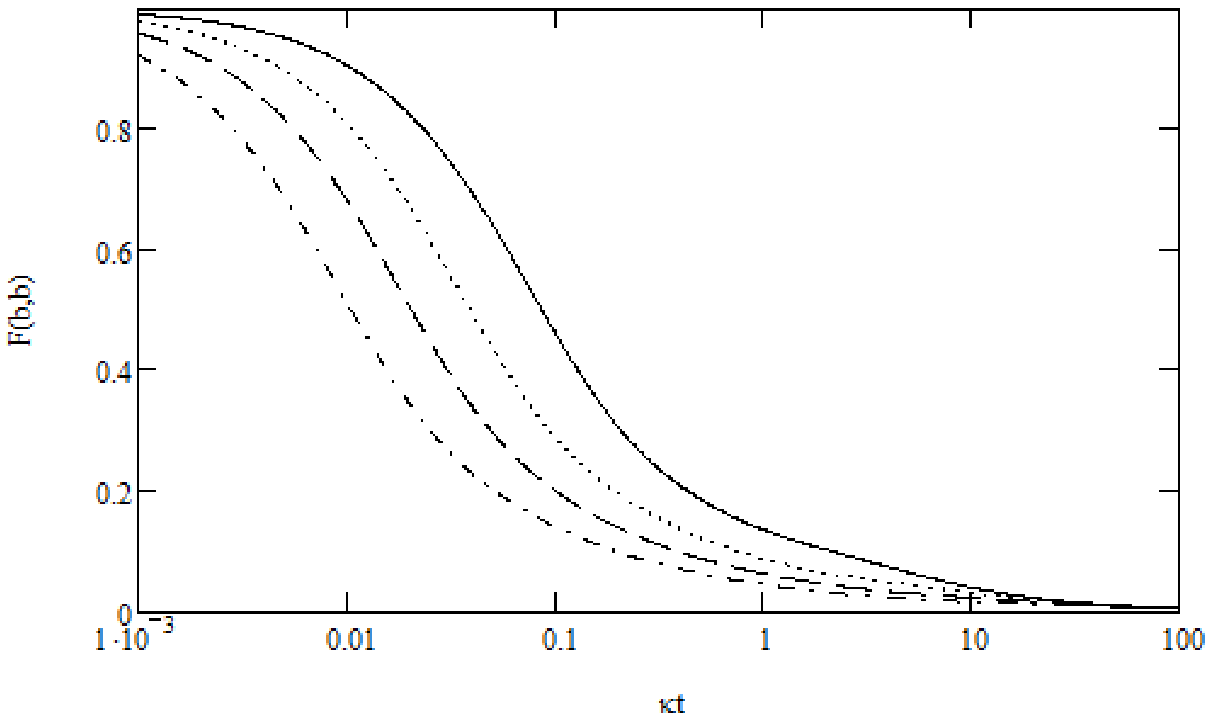} \vspace{-0cm}
\caption{Full thin line we see the fidelity of the state $|b><b|(t)$ and $%
|b> $ for $b=1$, dotted line $b=5$, dashed line $b=10$, dash dotted line $%
b=15$, the x axis corresponds to $\protect\kappa t$ in a log scale.}
\label{pb}
\end{figure}

In figure (\ref{pb}) we have the fidelity of the state $|b><b|(t)$ and $|b>$
for b=1, 5, 10 and 15. This is the probability of observing the state $|b>$
as function of time supposing we have prepared this state. As we can observe
in this figure, the higher energy gets, the faster fidelity approaches zero
(b). This means that we have a short time to measure the state before it
becomes a complete statistical mixture.

\bigskip If we consider a mixed state as (\ref{ro(t) para fock}) the $y(n)$
function has to be adapted. For the quartic oscilator que can assume that
\begin{equation*}
\left\langle \tau \right\rangle \approx \frac{2\pi \hbar \left\langle
\widehat{N}\right\rangle }{\left\langle \hat{H}_{0}\right\rangle },
\end{equation*}%
where $\widehat{N}=\hat{a}^{\dagger }\hat{a}.$ Thus, for an initial fock
state $\left\vert b\right\rangle ,$ we have

\begin{equation}
\left\langle \widehat{N}_{b}(b)\right\rangle =\sum_{l=0}^{\infty
}\sum_{p=0}^{b}\frac{b!(p+l)!(p+l)}{\left( p!\right) ^{2}!l!(b-p)!}\times
\gamma ^{b+l-p}(0,t)\zeta ^{2p+1}(0,t)  \label{Nmed}
\end{equation}%
and%
\begin{equation}
\left\langle \hat{H}_{0}(b)\right\rangle =\sum_{l=0}^{\infty }\sum_{p=0}^{b}%
\frac{b!(p+l)!\hbar \omega (p+l)+\lambda \hbar ^{2}(p+l)^{2}}{\left(
p!\right) ^{2}!l!(b-p)!}\times \gamma ^{b+l-p}(0,t)\zeta ^{2p+1}(0,t).
\label{H0med}
\end{equation}

\bigskip Then we have $\left\langle \Delta E_{n}\right\rangle =\left(
\left\langle \hat{H}_{0}(b)\right\rangle -\left\langle \hat{H}%
_{0}(b-1)\right\rangle \right) /2$ and $\left\langle \Delta \tau
_{b}\right\rangle =\left( \left\langle \tau _{b}\right\rangle -\left\langle
\tau _{b-1}\right\rangle \right) /2,$ and finally , for an intial state $%
\left\vert b\right\rangle $\ we obtain%
\begin{equation}
\left\langle y(b)\right\rangle =\left\langle \Delta E_{b}\right\rangle
\left\langle \Delta \tau _{b}\right\rangle .  \label{ynopen}
\end{equation}

\bigskip

\begin{figure}[th]
\vspace{-0cm} \includegraphics[scale=.8]{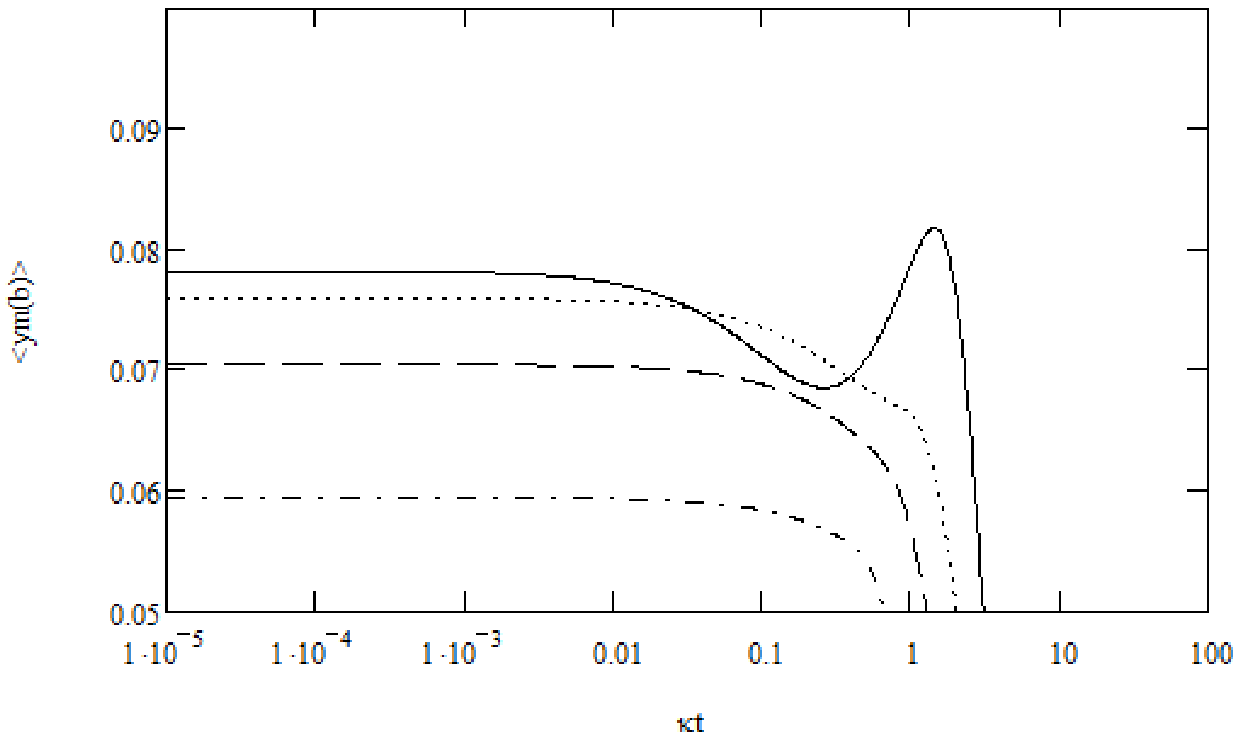} \vspace{-0cm}
\caption{Full thin line we see $\left\langle y(b)\right\rangle $ of state $%
|b><b|(t)$ for $b=2$, dotted line $b=5$, dashed line $b=10$, dash dotted
line $b=15$, the x axis corresponds to $\protect\kappa t$ in a log scale, $%
\protect\omega /\hbar \protect\lambda =0.10$.}
\label{ynm1}
\end{figure}

\begin{figure}[th]
\vspace{-0cm} \includegraphics[scale=.8]{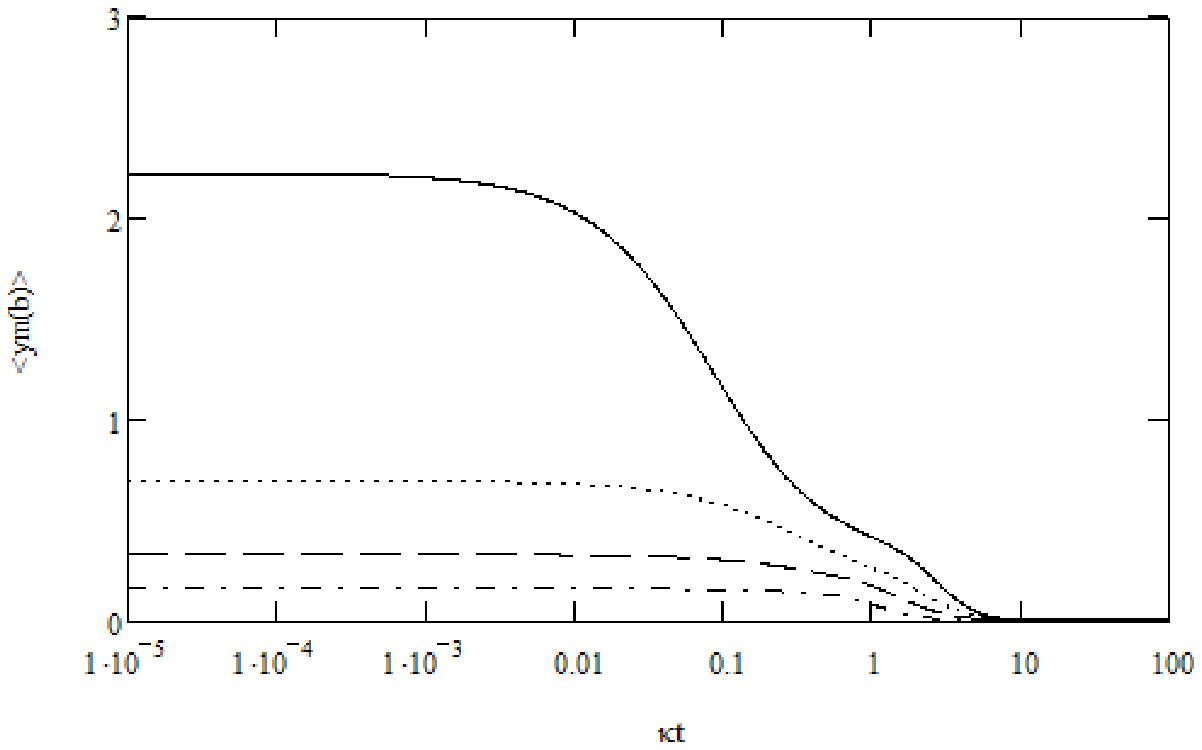} \vspace{-0cm}
\caption{Full thin line we see $\left\langle y(b)\right\rangle $ of state $%
|b><b|(t)$ for $b=2$, dotted line $b=5$, dashed line $b=10$, dash dotted
line $b=15$, the x axis corresponds to $\protect\kappa t$ in a log scale, $%
\protect\omega /\hbar \protect\lambda =10$.}
\label{ynm2}
\end{figure}

In figure (\ref{ynm1}) we have $\left\langle y(b)\right\rangle $ of the
state $|b><b|(t)$ for b=2, 5, 10 and 15 and $\omega /\hbar \lambda =0.10$.
In figure (\ref{ynm2}) we have $\left\langle y(b)\right\rangle $ of the
state $|b><b|(t)$ for b=2, 5, 10 and 15 and $\omega /\hbar \lambda =10$. As
we can see, the action of the environment reduces the chance of observing
the spectra discreteness even for hight non linearity. The the "lifetime" of
discreteness is approximately $1/\kappa$ , thus we connected the classical
limit of a spectra with the action of an environment and experimental
resolution.

\section{Conclusion}

In this work we have shown that the discrete nature of the energy levels can
be accessed by classical measurements in some cases. We also defined a
precise limit for this procedure using the function $y(n)=\left\vert \Delta
E_{n}\Delta \tau _{n}\right\vert $ and comparing it with the time-energy
uncertainty principle. This maneuver gives us a complementarity principle
and a well defined mathematical limit dictated by the experiment. Of course,
the fact that we are not able to recognize the discrete nature of a spectrum
does not necessarily mean it is not discrete. It only means how
\textquotedblleft myope\textquotedblright\ we are, suggesting that Classical
Mechanics can be viewed as a blurring of essential aspects of Quantum
Mechanics and also explains why it took so long to find quantum effects.
Also, as observed by many works \cite%
{Caldeira,Zur1996,Zur2003,Oliveira06,renato2006,Faria07,Wiebe,renato2007} a
quantum system is never isolated, thus we are forced to include environment
action that, in the present case, turns impractical any classical energy
measurement as defined in section III, what means that otherwise we show how
to get the classical limit in terms of discrete nature of the spectra.

\end{document}